\documentclass[pra,showpacs,twocolumn,typeset,superscriptaddress]{revtex4}

\usepackage{etoolbox} 
\usepackage{lipsum} 

\usepackage{amsmath}

\usepackage[capitalize]{cleveref}

\usepackage{amsfonts}

\usepackage{amssymb}
\usepackage{graphicx}
\usepackage{epsfig}
\usepackage{dcolumn}
\usepackage{bm}

\begin{document}
\preprint{APS}

\title{Threshold effect for probabilistic entanglement swapping}

\author{Luis Roa Oppliger}\email{lroa@udec.cl}
\affiliation{Departamento de F\'{\i}sica, Universidad de Concepci\'{o}n, Casilla 160-C, Concepci\'{o}n, Chile.}
\author{Torben L. Purz}
\affiliation{Faculty of Physics, Georg-August-Universit\"{a}t G\"{o}ttingen, G\"{o}ttingen, Germany}
\author{Ariana Mu\~noz}
\affiliation{Facultad de Ingenier\'{\i}a, Universidad Aut\'onoma de Chile, 5 Poniente 1670, Talca, Chile}
\author{Sebasti\'an Castro}
\affiliation{Departamento de F\'{\i}sica, Universidad de Concepci\'{o}n, Casilla 160-C, Concepci\'{o}n, Chile.}
\author{Gonzalo Hidalgo}
\affiliation{Departamento de F\'{\i}sica, Universidad de Concepci\'{o}n, Casilla 160-C, Concepci\'{o}n, Chile.}
\author{David Montoya}
\affiliation{Departamento de F\'{\i}sica, Universidad de Concepci\'{o}n, Casilla 160-C, Concepci\'{o}n, Chile.}
\date{\today}

\begin{abstract}
The basic \emph{entanglement swapping protocol} allows to project two qubits, which have never interacted, onto a maximally entangled state.
For deterministic swapping, the key ingredient is the maximal entanglement that was initially contained in two pairs of qubits and the capacity of projecting onto a Bell basis.
Thus the basic and deterministic entanglement swapping scheme involves three maximal level of entanglement.
In this work we propose probabilistic entanglement swapping processes performed with different amounts of initial entanglement.
Besides that we suggest a non Bell measuring-basis, to introduce a third entanglement level in the process.
Additionally, we propose the \emph{unambiguous state extraction scheme} as the local mechanism for probabilistically achieving the EPR projection.
The combination of these three elements allows us to design four strategies for performing probabilistic entanglement swapping.
Surprisingly, we find a twofold entanglement threshold effect related to the concurrence of the measuring-basis.
Specifically, the maximal probability of accomplishing a EPR projection becomes a constant for concurrences higher than or equal to threshold entanglement value.
Thus, we show that maximal entanglement in the measuring-basis is not required for attaining the EPR projection.
\end{abstract}

 \maketitle

\section{Introduction}

Erwin Schr\"odinger was one of the first to notice the existence of special correlation, entanglement,
present in a superposition of tensorial product states of two subsystems \cite{Schrodinger}.
Currently, it is known that entanglement is purely a quantum ingredient, which introduces non-local effects in protocols for processing information on atomic and molecular scales \cite{NielsenBook,APeres,Mermin,GAlber}.
Thus, a wide community of researchers have focused efforts to find a well defined functional which can assign entanglement values for pure and mixed states.
The most known of these functional is the entanglement of formation, which quantifies the resources needed to create a given entangled state \cite{CHBennett}.
Specifically, W. K. Wootters found a closed analytical formula for the concurrence of an arbitrary state of two qubits,
which is a monotone function of entanglement of formation \cite{Wootters1,Wootters2}.
In consequence, for two qubits, the concurrence can be used as a measurement for entanglement in its own merit,
which is what we consider here.
For a $2\otimes 2$ pure state, its concurrence can be evaluated at a glance.
If it is represented in the Schmidt decomposition; it is two times the product of its Schmidt coefficients \cite{Wootters1,Wootters2}.
So, a factorized state lacks concurrence, whereas a maximal entangled state (EPR) has a concurrence value equal to $1$.

In this context, the generation, manipulation, control, and the practical scope of this quantum correlation become important.
In particular, the capacity to distribute entanglement between distant systems has potential applications in designing innovative protocols of quantum information, whether or not they have a classical counterpart \cite{GBrassard,MZukowski,EKnill,HJBriegel,PEShor,VMKendon,MHillery,ASCoelho,JCRetamal,CJoshi,UAkram,PFacchi,CSpee,CasparG}.

Theoretically and experimentally the \emph{entanglement swapping}, as a mechanism for distributing the entanglement correlation, has been extensively studied \cite{MZukowski,BYurke,Yurke,Weinfurter,PKok,CBranciard,Adhikari,TieJunWang,JCho,AKhalique,SMRoy,GGruning,ShuChengLi,BTKirby,ChuanMeiXie,YHLiu,MNaseri}.

In this article, we analyze four schemes for performing probabilistic entanglement swapping, in which the main task is to project two qubits, which never have interacted, onto a Bell state.
We consider two pairs of qubits in different partially entangled pure states.
Besides, we also assume that the measuring-basis can consist of non-Bell states.
Additionally, we propose the \emph{unambiguous state extraction} (USE) protocol \cite{Hsu,BHe,Hutin,Gautam} as the mechanism for implementing local operations.

The article is organized as follows.
In Sec. \ref{sec1} we briefly recall the well known, deterministic and basic entanglement swapping scheme.
In Sec. \ref{sec2} we propose four strategies for achieving the task of probabilistically projecting onto an EPR state, shared by two qubits which never interacted.
In the last Sec. \ref{sec3} we summarize the principal results of this article.
Additionally, Appendix \ref{AppendixA} shows that the \emph{unambiguous state extraction} (USE) protocol \cite{Hsu,BHe,Hutin,Gautam} can be locally applied to
extract an EPR state with optimal probability.
In particular, we introduce an explicit joint unitary transformation, which allows to accomplish the EPR projection, additionally finding the optimal probability of success as a function of the initial concurrence value \cite{Bose,Vidal,Nielsen,BKraus,Guevara}.
The results described are directly applied onto the first, second and fourth strategy.

\section{Deterministic and basic entanglement swapping} \label{sec1}

\begin{figure}[h]
\center
\includegraphics[angle=360, width =0.4\textwidth]{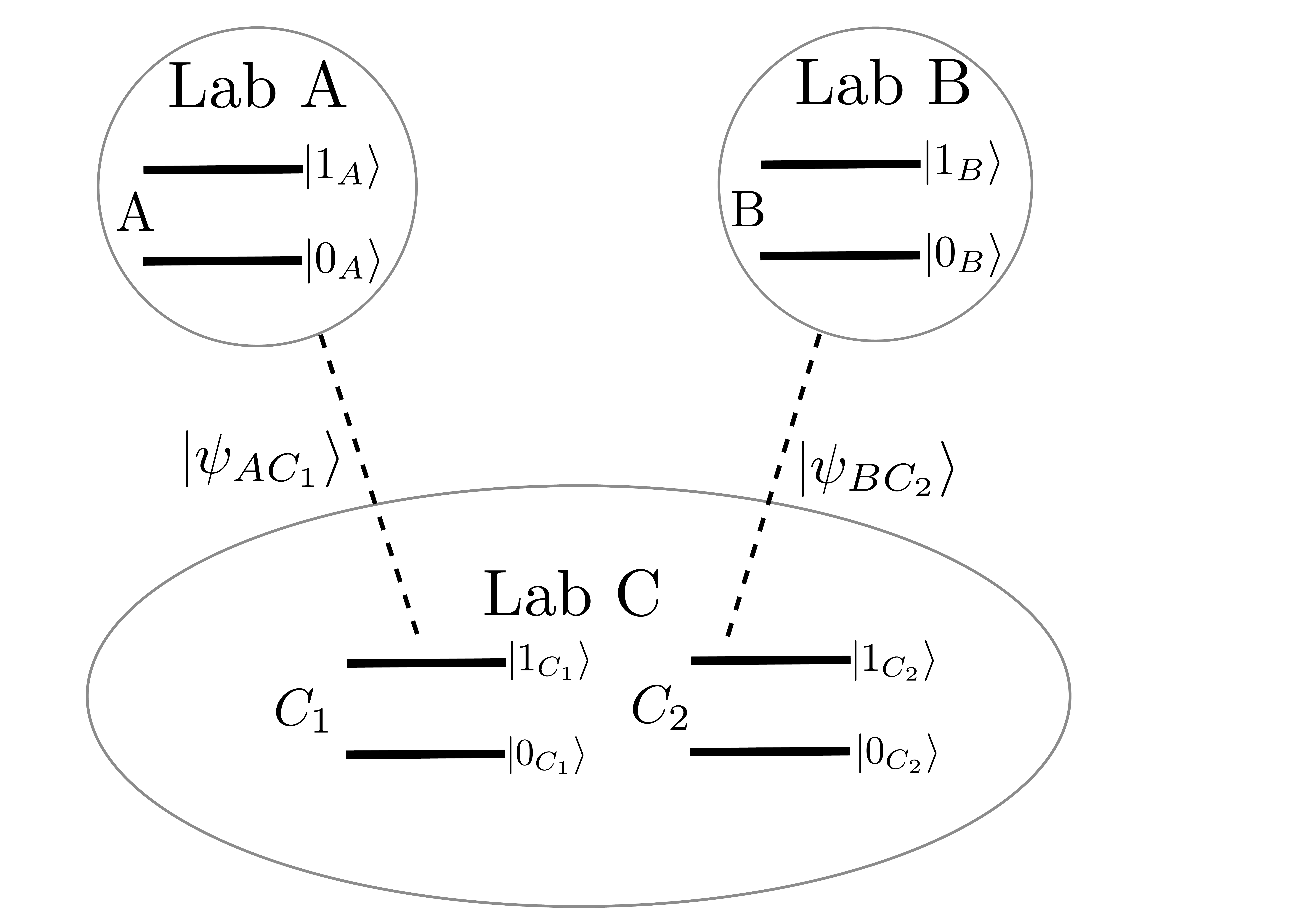}
\caption{Spatial distribution of the four qubits $A$, $B$, $C_1$ and $C_2$.
Qubits $C_1$ and $C_2$ are in the same laboratory.
The laboratories of the qubits $A$ and $B$ are away from each other and are also away from Lab-$C$.}
\label{figure1}
\end{figure}

The basic entanglement swapping scheme consists of four qubits $A$, $C_{1}$, $B$, $C_{2}$, the pairs $AC_{1}$ and $BC_{2}$ were prepared previously in Bell states.
The qubits $C_{1}$ and $C_{2}$ remain in the same laboratory (Lab-$C$), whereas the qubits $A$ and $B$ are taken to two different laboratories, away from each other and away from Lab-$C$.
Therefore, joint operations are only allowed between qubits $C_{1}$ and $C_{2}$.
Local operations can be applied onto each of the four qubits and classical communication can be enabled among all of them.
The purpose of this scheme is to entangle the pair $AB$ in an EPR state by using classical communication with local and joint operations.
Figure \ref{figure1} illustrates the spatial scheme of the different locations of four qubits.

The following identity gives account of the basic entanglement swapping procedure,
\begin{eqnarray}
|\psi _{AC_{1}}^{+}\rangle |\psi _{BC_{2}}^{+}\rangle &=&\frac{1}{2}\left(
|\phi _{C_{1}C_{2}}^{+}\rangle |\phi _{AB}^{+}\rangle  \!+|\psi _{C_{1}C_{2}}^{+}\rangle |\psi _{AB}^{+}\rangle  \right.   \nonumber \\
&&\!\left.\!-|\psi_{C_{1}C_{2}}^{-}\rangle |\psi _{AB}^{-}\rangle -|\phi_{C_{1}C_{2}}^{-}\rangle |\phi _{AB}^{-}\rangle  \! \right)\!,  \label{Ident1}%
\end{eqnarray}%
where,
\begin{eqnarray*}
|\psi ^{\pm }_{ij}\rangle &=& \left( |0_i\rangle |1_j\rangle \pm |1_i\rangle|0_j\rangle \right) /\sqrt{2}, \\
|\phi ^{\pm }_{ij}\rangle &=& \left( |0_i\rangle|0_j\rangle \pm |1_i\rangle |1_j\rangle \right) /\sqrt{2},
\end{eqnarray*}
are the Bell-basis for a pair of qubits $i\otimes j$ \cite{NielsenBook}.
The right hand side of Eq. (\ref{Ident1}) clearly shows the one-to-one correlation between the Bell states for the pairs $C_1C_2$ and $AB$.
From this, one realizes that by measuring the Bell states in Lab-$C$ the pair $AB$ can be projected onto a Bell state,
thus deterministically achieving the main task.
It is worth noting, that determinism demands the key ingredient of maximal entanglement in the two initial states and also in the measuring-basis.

\section{Probabilistic entanglement swapping}  \label{sec2}

We analyze four strategies for probabilistically obtaining an EPR state in the bipartite system $AB$, when the
pairs $AC_{1}$ and $BC_{2}$ initially are in the following partially entangled pure states,
\begin{subequations}
\begin{eqnarray}
|\bar{\psi}_{AC_{1}}\rangle &=&\alpha |0_{A}\rangle |1_{C_{1}}\rangle +\beta|1_{A}\rangle |0_{C_{1}}\rangle ,  \label{ps1} \\
|\bar{\psi}_{BC_{2}}\rangle &=&\gamma |0_{B}\rangle |1_{C_{2}}\rangle+\delta |1_{B}\rangle |0_{C_{2}}\rangle ,  \label{ps2}
\end{eqnarray}\label{ps}%
\end{subequations}%
where, without loss of generality, we assume the amplitudes $\alpha $, $\beta $, $\gamma $, $\delta $ to be real and non-negative numbers, such that,
\begin{equation}
\alpha \leq \beta, \text{ and }\gamma \leq \delta .  \label{wnlg}
\end{equation}
For normalization $\alpha ^{2}+\beta ^{2}=1$ and $\gamma ^{2}+\delta^{2}=1$.

The initial concurrences of the states (\ref{ps}) are $\mathcal{C}_{AC_{1}}=2\alpha \beta $ and $\mathcal{C}_{BC_{2}}=2\gamma \delta $, respectively.

\subsection{First strategy}

The first, simplest strategy is to extract an EPR state from each bipartite state (\ref{ps}) with the unitary-reduction local operations described in Appendix \ref{AppendixA}.

If both processes are successful, the basic swapping scheme can be carried out.
From Eqs. (\ref{Pext}) and (\ref{wnlg}) we get the success probability for obtaining an EPR state in the pair $AB$, which is equal to the product of the probabilities for extracting an EPR state from each state (\ref{ps}), specifically,
\begin{eqnarray}
P_{s_{1}} &=&4\alpha ^{2}\gamma ^{2}, \nonumber\\
&=&\left( 1-\sqrt{1-\mathcal{C}_{AC_{1}}^{2}}\right) \left( 1-\sqrt{1-\mathcal{C}_{BC_{2}}^{2}}\right). \label{Ps1}
\end{eqnarray}
Note, that $P_{s_{1}}$ is an increasing function of the initial concurrences, $\mathcal{C}_{AC_{1}}$ and $\mathcal{C}_{BC_{2}}$,
besides, entanglement values different from zero of both initial states are necessary and sufficient for having non zero success probability.

\subsection{Second strategy}

The second strategy is suggested by writing the initial tensorial product state $|\bar{\psi}_{AC1}\rangle|\bar{\psi}_{BC2}\rangle$ in the representation of Bell-basis of the bipartite system $C_1C_2$, i.e.,
\begin{eqnarray}
|\bar{\psi}_{AC1}\rangle|\bar{\psi}_{BC2}\rangle &=& \sqrt{\frac{p}{2}}|\phi_{C1C2}^{+}\rangle|\ddot{\phi}_{AB}^{+}\rangle \nonumber \\
&&+\sqrt{\frac{1-p}{2}}|\psi_{C1C2}^{+}\rangle|\ddot{\psi}_{AB}^{+}\rangle \nonumber \\
&&-\sqrt{\frac{1-p}{2}}|\psi _{C1C2}^{-}\rangle |\ddot{\psi}_{AB}^{-}\rangle  \nonumber \\
&&-\sqrt{\frac{p}{2}}|\phi _{C1C2}^{-}\rangle |\ddot{\phi}_{AB}^{-}\rangle,
\label{I2}
\end{eqnarray}
where we have defined the states for the pair $AB$ as follows,
\begin{subequations}
\begin{eqnarray}
|\ddot{\phi}_{AB}^{\pm }\rangle &=&\frac{\alpha \gamma |0\rangle |0\rangle
\pm \beta \delta |1\rangle |1\rangle }{\sqrt{p}},  \label{ssAB1} \\
|\ddot{\psi}_{AB}^{\pm }\rangle &=&\frac{\alpha \delta |0\rangle |1\rangle
\pm \beta \gamma |1\rangle |0\rangle }{\sqrt{1-p}},  \label{ssAB2}
\end{eqnarray}\label{ssAB}\end{subequations}%
and the probability,
\begin{equation*}
p=\alpha ^{2}\gamma ^{2}+\beta ^{2}\delta ^{2}.
\end{equation*}%
The identity (\ref{I2}) shows a one-to-one correlation between the Bell states for $C_1C_2$ and the states (\ref{ssAB}) for the pair $AB$.
Thus we realize that, by measuring Bell states in Lab-$C$, the bipartite system $AB$ is projected onto one of the partially entangled states (\ref{ssAB}) with probabilities $p/2$ and $(1-p)/2$, respectively.
Notice, that the minus sign in $|\ddot{\phi}_{AB}^{-}\rangle$ and $|\ddot{\psi}_{AB}^{-}\rangle$ can be removed by applying a local Pauli operator $\sigma _{z}$ on $A$ or $B$.

Therefore, in practice, the system $AB$ has two outcomes $|\ddot{\phi}_{AB}^{+}\rangle$ with probability $p$ and $|\ddot{\psi}_{AB}^{+}\rangle $ with probability $1-p$.

From each of these states, $|\ddot{\phi}_{AB}^{+}\rangle$ and $|\ddot{\psi}_{AB}^{+}\rangle$, one of the receivers, $A$ or $B$,
can probabilistically extract a Bell state by means of the local USE scheme.
If the outcome is $|\ddot{\phi}_{AB}^{+}\rangle $, the conditional probability of extracting a Bell state becomes $p_{ext,\phi}=2\alpha ^{2}\gamma ^{2}/p$,
otherwise, if the outcome is $|\ddot{\psi}_{AB}^{+}\rangle $, there are two cases for evaluating the conditional probability:\\
({\it i}) When $\alpha \delta \leq \beta \gamma $, which is equivalent to $\alpha \leq \gamma $, the conditional probability of successful extraction is $p_{ext,\psi }=2\alpha ^{2}\delta ^{2}/(1-p)$.\\
({\it ii}) When $\alpha \delta \geq \beta \gamma $, which is
equivalent to $\alpha\geq \gamma $, the conditional success probability is given by $p_{ext,\psi }=2\beta ^{2}\gamma ^{2}/(1-p)$.

\begin{figure}[h]
\center
\includegraphics[angle=360, width =0.4\textwidth]{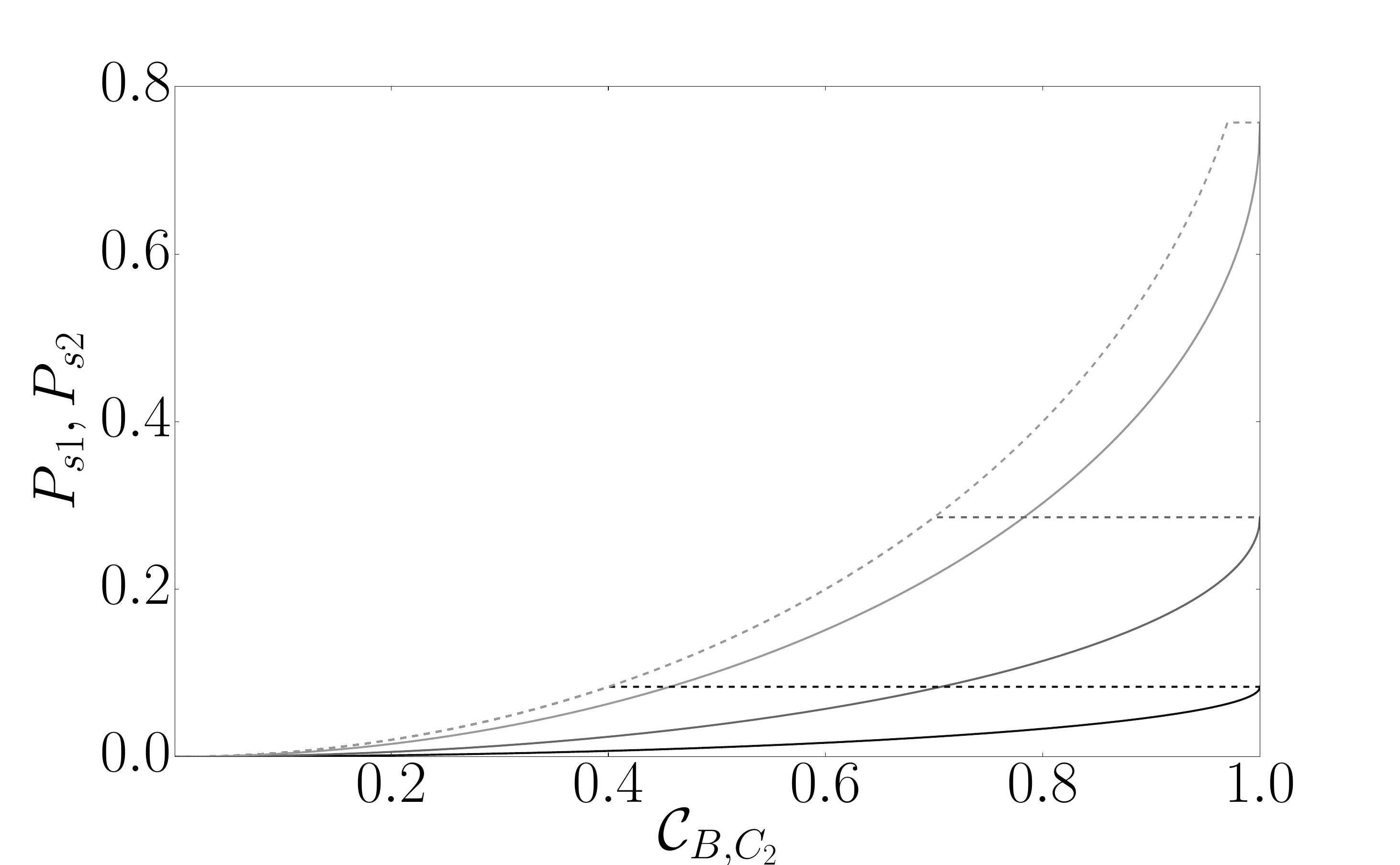}
\caption{Success probabilities for generating an EPR state as functions of $\mathcal{C}_{BC_{2}}$ with the first strategy (solid line) and with the second strategy (dashed line)
for different values of $\mathcal{C}_{AC_{1}}$: 0.4 (black), 0.7(gray), and 0.97 (light-gray).}
\label{figure2}
\end{figure}

Accordingly, the total success probability of achieving an EPR state in the pair $AB$ becomes,
\begin{eqnarray}
P_{s_{2}} &=&pp_{ext,\phi }+(1-p)p_{ext,\psi }, \nonumber \\
&=&2\min \left\{ \alpha ^{2},\gamma ^{2}\right\}, \nonumber \\
&=&\min \bigg\{1-\sqrt{1-\mathcal{C}_{AC_{1}}^{2}},1-\sqrt{1-\mathcal{C}_{BC_{2}}^{2}}\bigg\}. \label{Ps2}
\end{eqnarray}
Note, that the expression $P_{s_{2}}$ exhibits the entanglement threshold effect,
since it is determined by the smallest entanglement of the initial states $|\bar{\psi}_{AC_{1}}\rangle $ and
$|\bar{\psi}_{BC_{2}}\rangle $.
It is worth emphasizing, that both initial entanglements are necessary and sufficient for having a success probability different from zero.
Besides, by comparing the success probabilities of both strategies we find that,
\begin{equation}
P_{s_{2}}\geq P_{s_{1}}. \label{ineq1}
\end{equation}
The equality $P_{s_{2}}=P_{s_{1}}$ holds true, if at least one of the two initial concurrences is equal to $1$ or equal to zero.
For instance, by considering $\mathcal{C}_{AC_{1}}$ fixed, $P_{s_{2}}$ achieves its maximal value for $\mathcal{C}_{BC_{2}}=\mathcal{C}_{AC_{1}}$,
whereas $P_{s_{1}}$ reaches the same maximal value at $\mathcal{C}_{BC_{2}}=1$.
In other words, for a fixed value $\mathcal{C}_{AC_{1}}$, the maximal $P_{s_{2}}$ demands $\mathcal{C}_{BC_{2}}=\mathcal{C}_{AC_{1}}$, while the same maximal value for $P_{s_{1}}$ demands $\mathcal{C}_{BC_{2}}=1$.
Figure \ref{figure2} illustrates the behavior of the probabilities (\ref{Ps1}) and (\ref{Ps2}) for different values of $\mathcal{C}_{AC_{1}}$.

In consequence, if we consider entanglement as a resource, we can conclude that the second strategy is more efficient than the first one.

\subsection{Strategies without Bell-basis measurement}

Now, instead of projecting onto Bell states in $C_1C_2$, we propose to measure an observable that has the following eigenstates,
\begin{subequations}
\begin{eqnarray}
|\mu _{C_{1}C_{2}}^{+}\rangle &=&x|00\rangle +y|11\rangle ,  \label{b1} \\
|\mu _{C_{1}C_{2}}^{-}\rangle &=&y|00\rangle -x|11\rangle ,  \label{b2} \\
|\nu _{C_{1}C_{2}}^{+}\rangle &=&x|01\rangle +y|10\rangle ,  \label{b3} \\
|\nu _{C_{1}C_{2}}^{-}\rangle &=&y|01\rangle -x|10\rangle ,  \label{b4}
\end{eqnarray}\label{b}%
\end{subequations}%
where, without loss of generality, we can consider the amplitudes $x$ and $y$ to be real and non-negative numbers, and,
\begin{equation}
x\leq y.    \label{xmey}
\end{equation}
For normalization $x^2+y^2=1$.
Each state (\ref{b}) has the same concurrence $\mathcal{C}_{C_1C_2}(x)=2xy$, which is a function of the basis parameter $x$.

From here the natural questions to be addressed are:
\begin{itemize}
  \item Are there \emph{special values} for the concurrence $\mathcal{C}_{C_1C_2}(x)$, with regard to the initial ones $\mathcal{C}_{AC_{1}}$ and $\mathcal{C}_{BC_{2}}$,
for which the EPR projection can be accomplished with higher probability value?.
  \item Is a measurement on Bell states required to obtain the same value of probability reached with the second strategy?
\end{itemize}

We encountered two \emph{special values} for the concurrence $\mathcal{C}_{C_1C_2}(x)$, which are given by,
\begin{equation}
\mathfrak{C}_{C_1C_2}^\pm = \frac{\mathcal{C}_{AC_{1}}\mathcal{C}_{BC_{2}}}{1\pm\sqrt{(1-\mathcal{C}_{AC_{1}}^{2})(1-\mathcal{C}_{BC_{2}}^{2}})}. \label{Csvpm}
\end{equation}
In the search of them, we found two schemes, which are described below.

The two schemes are suggested by representing the initial state $|\bar{\psi}_{AC1}\rangle |\bar{\psi}_{BC2}\rangle$ in the basis (\ref{b}).
Specifically, we analyze the following identity,
\begin{eqnarray}
|\bar{\psi}_{AC1}\rangle |\bar{\psi}_{BC2}\rangle & = & \sqrt{p_{\mu ^{+}}}|\mu_{C_{1}C_{2}}^{+}\rangle |\ddot{\phi}_{AB}\rangle   \nonumber \\
&& +\sqrt{p_{\nu ^{+}}}|\nu_{C_{1}C_{2}}^{+}\rangle |\ddot{\psi}_{A,B}\rangle \nonumber \\
&& -\sqrt{p_{\nu ^{-}}}|\nu _{C_{1}C_{2}}^{-}\rangle |\dddot{\psi}_{A,B}\rangle \nonumber \\
&& -\sqrt{p_{\mu ^{-}}}|\mu _{C_{1}C_{2}}^{-}\rangle |\dddot{\phi}_{A,B}\rangle. \label{I3}
\end{eqnarray}
Here we have defined the possible outcome states of the qubits $AB$,
\begin{subequations}
\begin{eqnarray}
|\ddot{\phi}_{AB}\rangle &=&\frac{\alpha \gamma y|00\rangle +\beta \delta x|11\rangle }{\sqrt{p_{\mu ^{+}}}}, \label{outAB1}\\
|\ddot{\psi}_{AB}\rangle &=&\frac{\alpha \delta y|01\rangle +\beta \gamma x|10\rangle }{\sqrt{p_{\nu ^{+}}}}, \label{outAB2} \\
|\dddot{\psi}_{AB}\rangle &=&\frac{\alpha \delta x|01\rangle -\beta \gamma y|10\rangle }{\sqrt{p_{\nu ^{-}}}}, \label{outAB3} \\
|\dddot{\phi}_{AB}\rangle &=&\frac{\alpha \gamma x|00\rangle -\beta \delta y|11\rangle }{\sqrt{p_{\mu ^{-}}}}, \label{outAB4}
\end{eqnarray}
\label{outAB}
\end{subequations}
and their respective probabilities,
\begin{subequations}
\begin{eqnarray}
p_{\mu ^{+}} &=&\alpha ^{2}\gamma ^{2}y^{2}+\beta ^{2}\delta ^{2}x^{2}, \label{pwne1}\\
p_{\nu ^{+}} &=&\alpha ^{2}\delta ^{2}y^{2}+\beta ^{2}\gamma ^{2}x^{2}, \label{pwne2}\\
p_{\nu ^{-}} &=&\alpha ^{2}\delta ^{2}x^{2}+\beta ^{2}\gamma ^{2}y^{2}, \label{pwne3}\\
p_{\mu ^{-}} &=&\alpha ^{2}\gamma ^{2}x^{2}+\beta ^{2}\delta ^{2}y^{2}. \label{pwn4}
\end{eqnarray}
\label{pwne}
\end{subequations}
Here $x$ is an important measuring-basis parameter to be strategically set in each of the following schemes.
By this, we fix the required amount of entanglement of the measuring-basis (\ref{b}).

\subsubsection{Third strategy}

It is worth noting that the concurrences of the outcomes (\ref{outAB}) can attain the maximal value $1$, but at different $x$ values.
Specifically, the state $|\ddot{\phi}_{AB}\rangle$ is maximally entangled at $x=x_1$, the outcome $|\ddot{\psi}_{AB}\rangle$ at $x=x_2$,
the state $|\dddot{\psi}_{AB}\rangle$ at $x=x_3$, and $|\dddot{\phi}_{AB}\rangle$ at $x=x_4$, where,
\begin{subequations}\begin{eqnarray}
x_{1} &=&\frac{\alpha \gamma }{\sqrt{\alpha ^{2}\gamma ^{2}+\beta ^{2}\delta^{2}}}, \label{xsv1}\\
x_{2} &=&\frac{\alpha \delta }{\sqrt{\alpha ^{2}\delta ^{2}+\beta ^{2}\gamma^{2}}}, \label{xsv2}\\
x_{3} &=&\frac{\beta \gamma }{\sqrt{\alpha ^{2}\delta ^{2}+\beta ^{2}\gamma^{2}}},  \label{xsv3}\\
x_{4} &=&\frac{\beta \delta }{\sqrt{\alpha ^{2}\gamma ^{2}+\beta ^{2}\delta^{2}}}.  \label{xsv4}
\end{eqnarray}\label{xsv}\end{subequations}
In general, these $x_i$ are different and, according to (\ref{wnlg}), they are ordered as follows,
\begin{subequations}
\begin{eqnarray}
x_{1}\leq x_{2}\leq \frac{1}{\sqrt{2}}\leq x_{3}\leq x_{4},&&\quad \text{if } \alpha\leq \gamma, \label{xorder1}\\
x_{1}\leq x_{3}\leq \frac{1}{\sqrt{2}}\leq x_{2}\leq x_{4},&&\quad \text{if } \alpha\geq \gamma. \label{xorder2}
\end{eqnarray}\label{xorder}%
\end{subequations}%
This means that the possible EPR outcomes are displaced in the $x$ measuring-basis parameter.
Besides, by considering (\ref{wnlg}) with $\alpha\neq\beta$ and $\gamma\neq\delta$, we observe that $|\dddot{\phi}_{A,B}\rangle$ does not exhibit maximal entanglement, since $x_4\notin[0,1/\sqrt{2}]$ (see condition (\ref{xmey})).
On the other hand, the maximal entanglement outcome can be at $|\ddot{\psi}_{AB}\rangle$ or at $|\dddot{\psi}_{AB}\rangle$, depending on the relation between $\alpha$ and $\gamma$.
In addition, note that if (\ref{wnlg}) is not satisfied, then $|\dddot{\phi}_{A,B}\rangle$ exhibits maximal entanglement instead of $|\ddot{\phi}_{A,B}\rangle$.

Therefore, to obtain the EPR projection, we choose a $x_i$ associated with the greatest probability value.
By replacing the $x_i$ in their respective probability (\ref{pwne}), we realize that the greatest probability is $P_{s_3}$, which is given by,
\begin{eqnarray}
P_{s_3} &=&\left\{
\begin{tabular}{ll}
$p_{\nu ^{+}}(x_{2}),$ & if $\alpha \leq \gamma ,$ \\
$p_{\nu ^{-}}(x_{3}),$ & if $\alpha \geq \gamma ,$
\end{tabular}%
\right. \nonumber \\
& = & \frac{2\alpha ^{2}\beta ^{2}\delta ^{2}\gamma ^{2}}{\alpha ^{2}\delta^{2}+\beta ^{2}\gamma ^{2}}, \nonumber \\
& = & \frac{\mathcal{C}_{AC_1}\mathcal{C}_{BC_2}\mathfrak{C}_{C_1C_2}^-}{4}.   \label{pnu}
\end{eqnarray}
This probability is smaller than or equal to the ones found in the first and second strategy, i.e.,
\begin{equation}
P_{s_3}\leq P_{s_1}\leq P_{s_2}, \label{ineq2}
\end{equation}
thus, $P_{s_3}$ becomes a lower bound value for the probability of success.
Note also that $P_{s_3}\leq1/4$ and the equality only holds true for the limit of the basic scheme, i.e., $\mathcal{C}_{AC_1}=\mathcal{C}_{BC_2}=1$
and $x_1=x_2=x_3=x_4=1/\sqrt{2}$, which means that each outcome is maximally entangled and each one can be obtained with probability $1/4$.

In general, in spite of (\ref{ineq2}), it is worth realizing that $x_2$ and $x_3$ are different from $1/\sqrt{2}$, which means that maximal entanglement is not required in the measuring-basis (\ref{b}) in order to have a probability different from zero for obtaining an EPR projection in the outcome (\ref{outAB2}) or (\ref{outAB3}).
Specifically, the values $x_2$ and $x_3$ are associated with the same special value $\mathfrak{C}_{C_1C_2}^-$ of the concurrence,
\begin{equation*}
\mathfrak{C}_{C_1C_2}^-=\mathcal{C}_{C_1C_2}(x_2)=\mathcal{C}_{C_1C_2}(x_3).
\end{equation*}%
From Eq.(\ref{Csvpm}) we can note that if one of the two initial concurrences, $\mathcal{C}_{AC_1}$ or $\mathcal{C}_{BC_2}$, is equal to $1$, to say $\mathcal{C}_{AC_1}=1$, then $\mathfrak{C}_{C_1C_2}^-=\mathcal{C}_{BC_{2}}$.
If $\mathcal{C}_{AC_1}=\mathcal{C}_{BC_2}$, then we can conclude that $\mathfrak{C}_{C_1C_2}^-=1$.
On the other hand, if one of the two initial concurrences, $\mathcal{C}_{AC_{1}}$ or $\mathcal{C}_{BC_{2}}$, is equal to $0$, then $\mathfrak{C}_{C_1C_2}^-=0$ and $P_{s_3}=0$.

The significance of these results is that there occurs a probability different from zero of accomplishing the EPR projection without requiring maximal entanglement in the measuring-basis.
That motivates us to propose the following strategy.

Here we want to mention that in Ref. \cite{ChuanMeiXie} the authors focus on the study of behaviour and the relationship among the concurrences of the outcome states, the parameter of the measuring-basis, and the respective probabilities of the outcome states.

\subsubsection{Fourth strategy}

Once the measurement result (\ref{b}) is known, one of the receivers, $A$ or $B$, can probabilistically extract an EPR state by means of the local USE procedure from the respective outcome (\ref{outAB}).

According to Appendix \ref{AppendixA} the conditional probabilities of successfully extracton depend on the relation between $\alpha$ and $\gamma$ and is read as follows.
If the outcome is $|\ddot{\phi}_{AB}\rangle $, then the conditional probability $p_{ext,\ddot{\phi}}$ for extracting a Bell state becomes,
\begin{subequations}
\label{pext1234}
\begin{equation}
p_{ext,\ddot{\phi}}=\left\{
\begin{tabular}{ll}
$\frac{2\beta ^{2}\delta ^{2}x^{2}}{p_{\mu ^{+}}}$, & if $x\leq x_{1}$,\\
$\frac{2\alpha ^{2}\gamma ^{2}y^{2}}{p_{\mu ^{+}}}$, & if $x\geq x_{1}$.
\end{tabular}\right.
\label{pext1}
\end{equation}%
If the outcome is $|\ddot{\psi}_{A,B}\rangle $, then the conditional probability of success $p_{ext,\ddot{\psi}}$ is,
\begin{equation}
p_{ext,\ddot{\psi}}=\left\{
\begin{tabular}{ll}
$\frac{2\beta ^{2}\gamma ^{2}x^{2}}{p_{\nu ^{+}}}$, & if $x\leq x_{2}$, \\
$\frac{2\alpha ^{2}\delta ^{2}y^{2}}{p_{\nu ^{+}}},$ & if $x\geq x_{2}.$
\end{tabular}\right.
\label{pext2}
\end{equation}
If the outcome is $|\dddot{\psi}_{A,B}\rangle $, then the conditional success probability $p_{ext,\dddot{\psi}}$ reads,
\begin{equation}
p_{ext,\dddot{\psi}}=\left\{
\begin{tabular}{ll}
$\frac{2\alpha ^{2}\delta ^{2}x^{2}}{p_{\nu ^{-}}}$, & if $x\leq x_{3}$, \\
$\frac{2\beta ^{2}\gamma ^{2}y^{2}}{p_{\nu ^{-}}}$, & if $x\geq x_{3}$.
\end{tabular}\right.
\label{pext3}
\end{equation}
If the outcome is $|\dddot{\phi}_{A,B}\rangle $, then the conditional probability $p_{ext,\dddot{\phi}}$ of extracting an EPR state is given by,
\begin{equation}
p_{ext,\dddot{\phi}}=\frac{2\alpha ^{2}\gamma ^{2}x^{2}}{p_{\mu ^{-}}}. \label{pext4}
\end{equation}%
\end{subequations}%
Thus, from each possible outcome (\ref{outAB}) an EPR state can be probabilistically extracted.
Therefore, the total success probability is given by the sum of the probabilities (\ref{pwne}), each one multiplied by its respective conditional probability for extracting (\ref{pext1234}), i.e.,
\begin{equation*}
P_{s_{4}}(x)=p_{\mu ^{+}}p_{ext,\ddot{\phi}}+p_{\nu ^{+}}p_{ext,\ddot{\psi}}+p_{\nu ^{-}}p_{ext,\dddot{\psi}}+p_{\mu ^{-}}p_{ext,\dddot{\phi}}.
\end{equation*}%
Similarly, for evaluating $P_{s_{4}}(x)$, we must take into account the relation between $\alpha$ and $\gamma$.
In consequence, the total success probability becomes,
\begin{widetext}
\begin{equation}
P_{s_{4}}(x)=\left\{
\begin{tabular}{ll}
$2x^{2}$, & \quad if $x\leq x_{1}$, \\
$2\alpha ^{2}\gamma ^{2}+2\left( \alpha ^{2}\delta ^{2}+\beta ^{2}\gamma^{2}\right) x^{2}$, & \quad if $x_{1}\leq x\leq \min \left\{x_{2},x_{3}\right\}$, \\
$2\min \left\{ \alpha ^{2},\gamma ^{2}\right\}$, & \quad if $\min \left\{x_{2},x_{3}\right\} \leq x\leq 1/\sqrt{2}$.
\end{tabular}\right.
\label{Ps4}
\end{equation}
\end{widetext}
From the expression (\ref{Ps4}) we notice the following effects:
\begin{itemize}
\item The slope of $P_{s_{4}}(x)$ exhibits two discontinuities, at $x_{1}$ and $\min \left\{ x_{2},x_{3}\right\}$.
\item The probability increases for $x\in [0,\min \left\{x_{2},x_{3}\right\}]$,
hence for $x\in [ \min \left\{x_{2},x_{3}\right\},1/\sqrt{2}]$ the probability $P_{s_{4}}(x)$ is constant and is equal to the maximal value obtained in the second strategy.
\item The threshold entanglement value is found, equal to the threshold value of the second strategy, but here the threshold value depends on $\mathcal{C}_{AC_{1}}$ and $\mathcal{C}_{BC_{2}}$ for all $x\in [ \min \left\{x_{2},x_{3}\right\},1/\sqrt{2}]$.
\item The maximal probability $2\min \left\{ \alpha^{2},\gamma ^{2}\right\}$ is achieved at $x=\min \left\{ x_{2},x_{3}\right\}$, which in general is smaller than $1/\sqrt{2}$.
      Therefore, the maximal entanglement in the measuring-basis (\ref{b}) is not required for reaching the optimal success probability.
\item There are two \emph{special values}, $\mathfrak{C}_{C_1C_2}^\pm$, for measuring-basis concurrence, in which $P_{s_{4}}(x)$ changes its behavior;
      at $\mathcal{C}_{C_1C_2}=\mathfrak{C}_{C_1C_2}^\pm$ the probability (\ref{Ps4}) \emph{abruptly changes} its slope,
      but for all $\mathcal{C}_{C_1C_2}(x)\geq\mathfrak{C}_{sv-}$ the probability $P_{s_{4}}(x)$ remains constant at its maximum value.
\end{itemize}
Although $\mathfrak{C}_{C_1C_2}^-$ is different from $\mathcal{C}_{AC_{1}}$ and $\mathcal{C}_{BC_{2}}$, it plays the role of an \emph{entanglement threshold value}, which is a function of the initial concurrences.

\begin{figure}[h]
\center
\includegraphics[angle=360, width =0.4\textwidth]{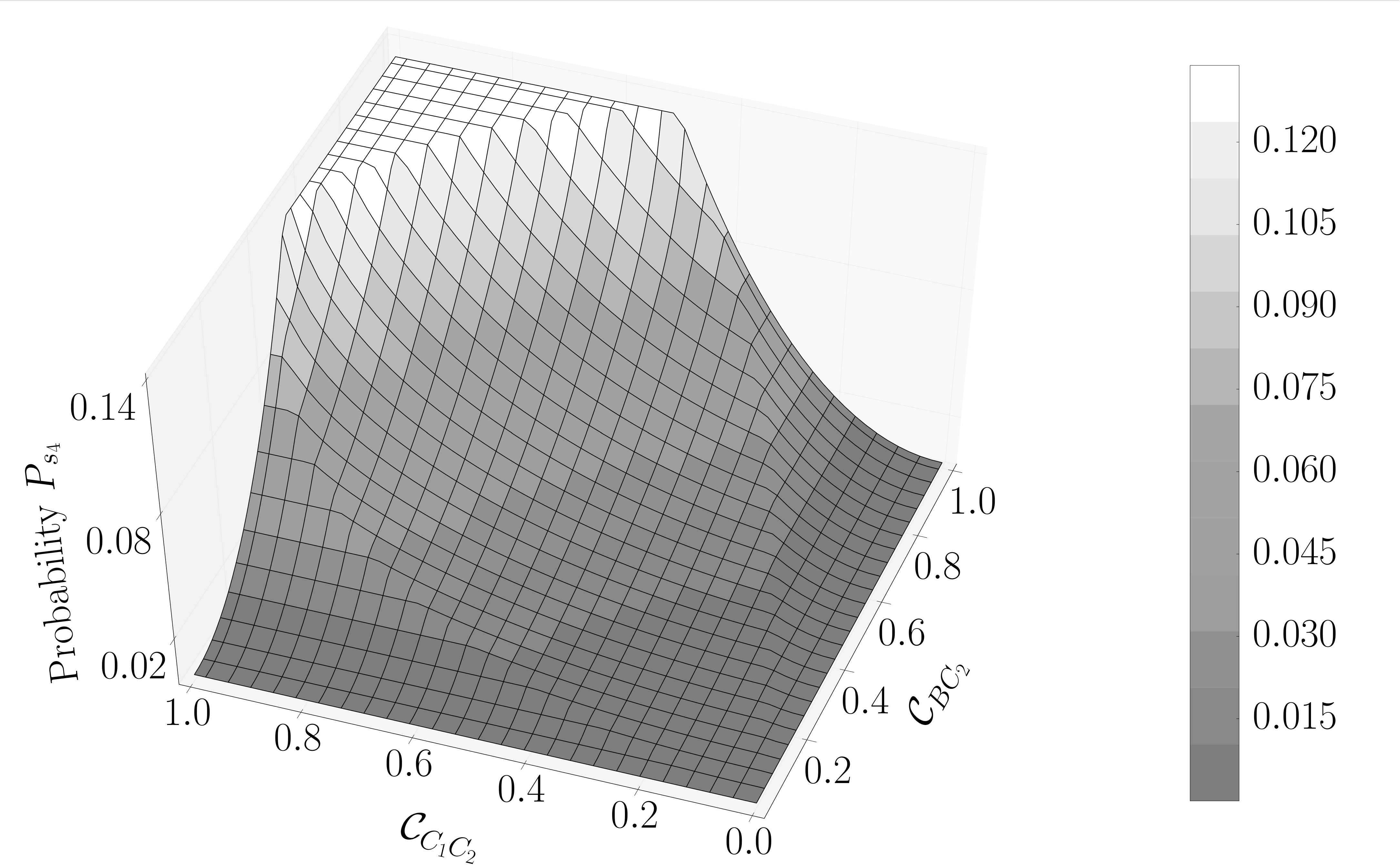}
\caption{Success probability $P_{s_{4}}$ as a function of $\mathcal{C}_{BC_{2}}$ and $\mathcal{C}_{C_1C_2}$ for $\mathcal{C}_{AC_{1}}=0.7$.}
\label{figure3}
\end{figure}

This strategy combines the best characteristics of the second and third strategy, i.e.,
it has the highest probability, obtained in the second strategy, and demands the special value for the concurrence of the measuring-basis, found for the third strategy.
Figure \ref{figure3} illustrates the behavior of (\ref{Ps4}) as a function of $\mathcal{C}_{BC_{2}}$ and $\mathcal{C}_{C_1C_2}$ for $\mathcal{C}_{AC_{1}}=0.7$; the increasing surface clearly shows the \emph{abrupt change} of slope, and the plateau in the surface corresponding to the twofold threshold effect.

Therefore, by considering the greatest probability of success and the entanglement as a resource, the fourth strategy is more efficient than the others three.

\section{Conclusions} \label{sec3}

We have addressed an unambiguous entanglement swapping scheme, with the main task of projecting two qubits onto an EPR state.
In our analysis, the target is to maximize the probability of success and to minimize the required entanglement in the measuring-basis.

We have introduced three levels of entanglement, two of them in the initial states and another one by considering non-Bell state in the measuring-basis.
Additionally, we proposed the \emph{unambiguous state extraction} scheme as mechanism for implementing the unitary-reduction local operator.

These considerations allow us to design four strategies for achieving the EPR projection.
The first one enables to accomplish the EPR projection, but with non-optimal probability of success.
For the second strategy, we found the upper bound value of the success probability, related to an entanglement threshold effect between the two concurrences of the initial states, i.e.,
the maximal success probability is determined by the smallest entanglement value of the initial states.
The third strategy has the lower bound value of the success probability, but gives account that the main task can be accomplished without maximal entanglement in the measuring-basis.
Thus, the three schemes lead us to suggest the fourth strategy.
We consider the fourth strategy as the optimal one, because it combines the best characteristics of the second and third strategy, i.e.,
it performed with the upper bound value of the success probability, found also in the second strategy, and it demands the non-maximal \emph{special value} for the concurrence of the measuring-basis, found also for the third strategy.
We realized that the concurrences special value $\mathfrak{C}_{C_1C_2}^-$ plays the role of a threshold effect between it and the initial concurrences.
Therefore, there is a scheme with optimal success probability, which does not require maximal entanglement in the measurement process.
In other words, the entanglement threshold effect between the two initial concurrences matches with $\mathfrak{C}_{C_1C_2}^-$;
thus, any concurrence value greater than $\mathfrak{C}_{C_1C_2}^-$ does not affect the success probability, i.e., it remains constant.

Besides, we found another special value $\mathfrak{C}_{C_1C_2}^+$, for which the increasing success probability changes its slope abruptly.

Here we have shown that, in order to accomplish the main task of the entanglement swapping with maximal probability, the maximal entanglement in measuring-basis is not required, equivalently, the \emph{special value} $\mathfrak{C}_{C_1C_2}^-$ is necessary and sufficient.

\appendix

\section{Optimal EPR extraction} \label{AppendixA}

Here we succinctly show that the locally applied USE protocol \cite{Hsu,Hutin} allows us to extract an EPR state, with optimal probability, from a partially entangled
state $|\bar{\psi}_{ij}\rangle $, shared by the qubits $i$ and $j$, i.e.,
\begin{equation}
|\bar{\psi}_{ij}\rangle \overset{\text{local USE}}{\rightarrow }\frac{|0_{i}\rangle |1_{j}\rangle +|1_{i}\rangle |0_{j}\rangle }{\sqrt{2}},
\label{Eext}%
\end{equation}%
with,
\begin{equation}
|\bar{\psi}_{ij}\rangle =u|0_{i}\rangle |1_{j}\rangle +v|1_{i}\rangle|0_{j}\rangle .  \label{ieA}
\end{equation}%
The $\{|0\rangle ,|1\rangle \}$ are eigenstates of the Pauli operator $\sigma _{z}$ of the qubit labeled by the subindex.
Without loss of generality, we assume that $u$ and $v$ are real and non-negative numbers and due to normalization $u^{2}+v^{2}=1$.
The entanglement of the initial state (\ref{ieA}) can be valued by the concurrence $\mathcal{C}_{\bar{\psi}}=2uv$ \cite{Wootters1,Wootters2}.

The local operation can be indistinctly applied onto qubit $i$ or $j$.
For instance, let us consider the qubit $i$ and an auxiliary qubit $a$, initially in the state $|0_{a}\rangle $.
Because the probability amplitude must have a module smaller than or equal to $1$, we have to consider two cases:\\
({\it i}) If $u\leq v$, we apply the joint unitary $U_{ia}$ onto the tensorial product state $|\bar{\psi}_{ij}\rangle |0_{a}\rangle $, with,
\begin{eqnarray}
U_{ia} &=&|0_{i}\rangle \langle 0_{i}|\otimes I_{a}+|1_{i}\rangle \langle 1_{i}|\otimes U_{a}, \label{uia1}\\
U_{a}|0_{a}\rangle  &=&\frac{u}{v}|0_{a}\rangle +\sqrt{1-\frac{u^{2}}{v^{2}}}|1_{a}\rangle , \nonumber
\end{eqnarray}%
where $I_{a}$ is the identity operator of the auxiliary qubit $a$.
Thus, $U_{ia}$ transforms $|\bar{\psi}_{ij}\rangle |0_{a}\rangle $ as follows,
\begin{eqnarray*}
U_{ia}|\bar{\psi}_{ij}\rangle |0_{a}\rangle  &=&\sqrt{2}u\frac{|0_{i}\rangle |1_{j}\rangle +|1_{i}\rangle |0_{j}\rangle }{\sqrt{2}}|0_{a}\rangle  \\
&&+\sqrt{v^{2}-u^{2}}|1_{i}\rangle |0_{j}\rangle |1_{a}\rangle ,
\end{eqnarray*}%
from where we realize, that by measuring $\sigma _{z}$ of the auxiliary qubit, the pair $ij$ is projected onto an EPR state with probability $p_{ext}=2u^{2}$, otherwise the correlation is lost.\\
({\it ii}) If $u\geq v$, we must apply the following joint unitary,
\begin{eqnarray}
U_{ia} &=&|0_{i}\rangle \langle 0_{i}|\otimes U_{a}+|1_{i}\rangle \langle 1_{i}|\otimes I_{a}, \label{uia2} \\
U_{a}|0_{a}\rangle  &=& \frac{v}{u}|0_{a}\rangle +\sqrt{1-\frac{v^{2}}{u^{2}}}|1_{a}\rangle,   \nonumber
\end{eqnarray}%
to obtain,
\begin{eqnarray*}
U_{ia}|\bar{\psi}_{ij}\rangle |0_{a}\rangle  &=&\sqrt{2}v\frac{|0_{i}\rangle |1_{j}\rangle +|1_{i}\rangle |0_{j}\rangle }{\sqrt{2}}|0_{a}\rangle  \\
&&+\sqrt{u^{2}-v^{2}}|0_{i}\rangle |1_{j}\rangle |1_{a}\rangle.
\end{eqnarray*}%
Similarly, by measuring $\sigma _{z}$ of the auxiliary qubit, the pair $ij$ is projected onto an EPR state with probability $p_{ext}=2v^{2}$,
otherwise the correlation is lost.

Therefore, for any given bipartite pure state $|\bar{\psi}_{ij}\rangle $ the probability $p_{ext}$ of extracting an EPR state by means of local operators and one way classical communication becomes,
\begin{eqnarray}
p_{ext}&=&2\min \left\{ u^{2},v^{2}\right\} ,  \nonumber \\
&=&1-\sqrt{1-\mathcal{C}_{\bar{\psi}}^2},                 \label{Pext}
\end{eqnarray}%
where $u$ and $v$ are its Schmidt coefficients.

We highlight that the probability expression(\ref{Pext}) agrees with the optimal ones found in Refs. \cite{Bose,Vidal,Nielsen,Guevara}.
It is worth mentioning here that, the joint unitaries (\ref{uia1}) and (\ref{uia2}) can be implemented experimentally in different physical systems \cite{JICirac,FDMartini,LIsenhower,KMMaller}
by composition of local unitaries and CNOT gates \cite{TSleator,DPDiVincenzo,VBuzek}.

Additionally, if the initial state is $|\bar{\phi}_{ij}\rangle =u|0_{i}\rangle |0_{j}\rangle +v|1_{i}\rangle|1_{j}\rangle$, then it can be transformed to $|\bar{\psi}_{ij}\rangle$
by applying the local unitary rotation $U_j=e^{-i(I-\sigma_x)pi/2}$ onto qubit $j$, and the above described scheme can be implemented.

\begin{acknowledgements}
\noindent Authors thank grants FONDECyT 1161631, and DAAD RISE Worldwide CL-PH-1736.
\end{acknowledgements}

\end{document}